\documentclass[superscriptaddress,twocolumn,amsmath,amssymb,amsthm]{revtex4}
\usepackage{amsmath}
\usepackage{graphicx}
\usepackage{epstopdf}
\usepackage{float}
\usepackage{hyperref}
\usepackage{subfigure}
\usepackage{color}
\usepackage[T1]{fontenc}
\usepackage[utf8]{inputenc}
\usepackage[toc,page]{appendix}
\usepackage[usenames,dvipsnames]{xcolor}
\usepackage[normalem]{ulem}
\usepackage{amssymb}
\usepackage{mathrsfs}
\usepackage{grffile}

\begin{document}

\title{Superfluid $\lambda$ Transition in Charged AdS Black Holes}

\author{Ning-Chen Bai}
\email{bainingchen@stu.scu.edu.cn}
\affiliation{Center for Theoretical Physics, College of Physics, Sichuan University, Chengdu, 610065, China}

\author{Lei Li}
\email{lilei78@ustc.edu.cn}
\affiliation{School of Physics and Electronic Engineering, \\ Sichuan University of Science and Engineering, Yibin, 643000, China}
 
\author{Jun Tao}
\email{taojun@scu.edu.cn}
\affiliation{Center for Theoretical Physics, College of Physics, Sichuan University, Chengdu, 610065, China}

\begin{abstract}
We observe a superfluid $\lambda$ transition in the $P-V$ criticality of charged AdS black holes, within a holographic extended thermodynamics that considers the variation of Newton's constant $G$. We calculate the critical
exponents and find that they coincide with those of a superfluid transition in liquid $^4\text{He}$ and the Bose-Einstein condensation of hard-sphere Bose gas. Moreover, the independence of entropy and thermodynamic volume in the holographic framework allows us to construct a well-defined Ruppeiner metric. The associated scalar curvature suggests that the black holes show similar microscopic interactions with the hard-sphere Bose gas, where the superfluid (condensed) phase is dominated by repulsive interactions, while the normal (gas) phase is dominated by attractive interactions. These findings might provide us with new insights into the quantum aspect of charged AdS black holes.
\end{abstract}

\maketitle
\newpage 

Thermodynamics continues to be one of the most fascinating areas in the study of black hole physics. The history of research began with the establishment of Hawking temperature and Bekenstein-Hawking
entropy for black holes, i.e., \cite{Hawking:1975vcx,Bekenstein:1972tm,Bekenstein:1973ur}
\begin{equation}
T=\frac{\hbar \kappa}{2 \pi c k_{B}}, \quad S=\frac{k_{B} c^{3} A}{4 \hbar G}, \label{eq:T_S}
\end{equation}
where $\kappa$ is the surface gravity and $A$ denotes the horizon area. Interestingly enough, these equations relates fundamental constants from all areas of modern physics: statistical mechanics,
gravity, and quantum mechanics. Such a feature indicates that black hole thermodynamics plays an essential role in understanding the nature of quantum gravity.

Over the past few years, there has been a notable emphasis on studying the thermodynamic of anti–de Sitter (AdS) black holes within an extended phase space, where the negative cosmological constant $\Lambda$ is considered to be a variable thermodynamic pressure \cite{Caldarelli:1999xj,Kastor:2009wy,Dolan:2010ha}
\begin{equation}
P=-\frac{\Lambda}{8 \pi G}, \quad \text{with} \quad \Lambda=- \frac{(D-1) (D-2)}{2 L^2}, \label{eq:P}   
\end{equation}
where $L$ is the AdS radius and $G$ represents the $D$-dimensional Newton's constant. This identification introduces an additional $V \delta P$ term in the first law of black hole thermodynamics, with $V$ representing the thermodynamic volume, allowing us to study the $P-V$ criticality of AdS black holes. This enables a comparison of the ``right” physical quantities between charged AdS black holes and van der Waals (VdW) fluids, highlighting the striking similarity between their respective phase transitions \cite{Chamblin:1999tk,Chamblin:1999hg,Kubiznak:2012wp}. Plenty of new phase behaviors have also been discovered in this field, such as the reentrant
phase transition \cite{Gunasekaran:2012dq,Altamirano:2013ane}, triple points \cite{Altamirano:2013uqa, Wei:2014hba}, superfluid transitions \cite{Hennigar:2016xwd,Hennigar:2016ekz}, etc. These findings further confirm the similarity between the phase transitions of AdS black holes and those of ordinary thermodynamic systems.

Nevertheless, understanding the holographic interpretation of this extended thermodynamics is somewhat challenging \cite{Kastor:2014dra,Karch:2015rpa,Johnson:2014yja,Dolan:2014cja,Zhang:2014uoa,Zhang:2015ova,Dolan:2016jjc,McCarthy:2017amh,Ahmed:2023snm,Frassino:2022zaz}. According to the AdS/CFT correspondence \cite{Maldacena:1997re,Gubser:1998bc,Witten:1998qj}, it has been argued that varying the cosmological constant $\Lambda$ is equivalent to varying the central charge $C$ or the number of colors $N$ in the dual CFT \cite{Kastor:2009wy,Kastor:2014dra,Karch:2015rpa,Johnson:2014yja,Dolan:2014cja}, via a holographic dictionary
\begin{equation}
C=k \frac{L^{D-2}}{16 \pi G}, \label{eq:central}  
\end{equation}
where $k$ is a numerical factor determined by specific features of the holographic system \cite{Karch:2015rpa}. As a result, the variation of central charge $C$ is fully dependent on that of the CFT volume $\mathcal{V} \sim L^{D-2}$ (supposing the curvature radius coincides with the AdS radius), rendering the CFT first law of the extended thermodynamics somewhat ambiguous. 

This issue can be addressed by including the variation of Newton’s constant $G$ in the bulk first law \cite{Visser:2021eqk,Cong:2021fnf}. It is then possible to fix the central charge $C$ so that the field theory remains unchanged, and varying $\Lambda$ will alter the CFT volume $\mathcal{V}$ more naturally. To be specific, for electrically charged Reissner--Nordstr\"{o}m AdS black holes, the complete bulk first law can be expressed as (setting $\hbar=c=k_B=1$ hereafter)
\begin{align}
\delta M &= \frac{\kappa}{8 \pi G} \delta A + \Phi \delta Q - \frac{V}{8 \pi G} \delta \Lambda -\left(M - \frac{\Phi Q}{2} \right) \frac{\delta G}{G}, \label{eq:bulk_first}
\end{align}
where $M$, $Q$ and $\Phi$ are the mass, electric charge and electric potential of the black hole, respectively. Further inserting the duality relations for the energy, electric charge and electric potential of CFT, Eq. (\ref{eq:bulk_first}) can be exactly converted to the CFT first law \cite{Visser:2021eqk}.

It is worth noting that the last term in the bulk first law does not directly correspond to a  thermodynamic interpretation. To obtain a first law with a well-defined thermodynamic correspondence, one can substitute \cite{Cong:2021fnf}
\begin{equation}
\frac{\delta G}{G}=-\frac{2}{D} \frac{\delta C}{C}-\frac{D-2}{D} \frac{\delta P}{P}
\label{eq:delta_G}
\end{equation}
into Eq. (\ref{eq:bulk_first}). The result shows that
\begin{equation}
\delta M=T \delta S+\Phi \delta Q +V_C \delta P+\mu \delta C, \label{eq:first_mixed}
\end{equation}
where 
\begin{equation}
V_C=\frac{2M + (D-4) \Phi Q}{2 D P}, \quad \mu=\frac{2 P(V_C- V )}{C(D-2)}
\end{equation}
are the new thermodynamic volume and chemical
potential, respectively. In the followings, we drop the $C$ in $V_C$ and reassign the new thermodynamic volume as $V$. On the basis of this mixed form of first law, the phase transition
of charged AdS black holes was reconsidered \cite{Cong:2021fnf}, and a VdW-like behaviour governed by the central charge $C$ was found, providing us with a new understanding of the critical behaviour of charged AdS black holes within the framework of gauge-gravity duality.

In this work, we reconsider the $P-V$ criticality of charged AdS black holes in the context of the mixed first law (\ref{eq:first_mixed}). Interestingly, as we shall see, instead of the VdW-like behaviour, the black holes exhibit a $\lambda$-line phase transition reminiscent of the superfluid transition in liquid $^4\text{He}$. Since the superfluidity is the direct consequence of quantum statistics effects \cite{Pathria:1996hda,huang1987}, this phenomenon may reveal more
features about the quantum aspect for the underlying microscopic degrees of freedom of the black holes.

We begin with the introduction of the gravity model. The action describing $(3+1)$-dimensional charged AdS black holes is given by
\begin{equation}
I =  \frac{1}{16\pi G} \int d^{4} x  \sqrt{-g}\left(R-2\Lambda - G F^{2}\right),
\label{eq:action}
\end{equation}
where $F^2=F_{\mu \nu} F^{\mu \nu}$, $F_{\mu \nu}=\partial_{\mu} A_{\nu}-\partial_{\nu} A_{\mu}$ is the electrodynamics field tensor and $A_{\mu}$ the gauge potential. Here we adopt the standard convention in the study of gravitational physics, in which case the stress-energy part of Einstein equations vanishes when $G \rightarrow 0$. This also leads to the factor $1/2$ in front of the $\Phi Q$ term in Eq. (\ref{eq:bulk_first}).

The electrically charged Reissner--Nordstr\"{o}m AdS black hole solution is given by
\begin{align}
d s^{2} &= -f(r) d t^{2}+ \frac{1}{f(r)} d r^{2}+r^{2}d\Omega_{2}^{2}, \nonumber \\
f(r) & =1-\frac{2 G M}{r}+\frac{G Q^2}{r^{2}} +\frac{r^{2}}{L^{2}},
\label{metric}
\end{align}
where $d\Omega_{2}^{2}$ denotes the line element of unit $2$-sphere. The associated mass, electromagnetic potential, entropy and temperature of the black hole are calculated as
\begin{align}
M&=\frac{r_+}{2 G} \left(1+\frac{r_+^{2}}{L^{2}}+\frac{G Q^2}{r_+^{2}} \right), \label{eq:mass}\\
\Phi &=\frac{Q}{r_{+}}, \quad S =\frac{\pi r_{+}^{2}}{G}, \label{eq:entropy} \\ 
T &= \frac{1}{4 \pi r_+}\left(1+ \frac{3 r_{+}^2}{L^{2}} -\frac{G Q^2}{r_+^{2}}\right),
\label{eq:tem}
\end{align}
where $r_+$ is the horizon radius determined as the largest root of $f(r_+)=0$. 

Utilizing the definition of bulk pressure Eq. (\ref{eq:P}) and CFT central charge Eq. (\ref{eq:central}), one can transform the above expressions for $(L, G)$ to $(P, C)$, and for simplicity we take $k=16 \pi$ in the followings. In the $D=4$ case,
\begin{equation}
L= \left(\frac{3 C}{8 \pi P}\right)^{1/4}, \quad G= \left(\frac{3}{8 \pi P C}\right)^{1/2}, \label{eq:G}
\end{equation}
then the black hole entropy and temperature from Eqs. (\ref{eq:entropy}) and (\ref{eq:tem}) can be written as
\begin{align}
S(r_+,C,z^i) & =\pi r_+^2 \sqrt{C/x}, \label{eq:S} \\
T(r_+,C,z^i) &=\frac{1}{4 \pi  r_+} \left(1+ \frac{ 3 r_+^2}{\sqrt{x C}} -\frac{Q^2}{r_+^{2}} \sqrt{\frac{x}{C}}\right), \label{eq:T} 
\end{align}
where $z^i=(Q,P)$ are the bulk parameters and $x=\frac{3}{8 \pi P}$. The thermodynamic volume from Eqs. (\ref{eq:first_mixed}) and (\ref{eq:mass}) is given by
\begin{align}
V(r_+,C,z^i) &=\frac{\pi}{3} \left( \sqrt{x C} r_+ +  r_+^3 + \frac{ x Q^2}{r_+}\right). \label{eq:V}
\end{align}

\begin{figure*}
\center{\subfigure{\label{Superfluid_P_V}
\includegraphics[height=5cm]{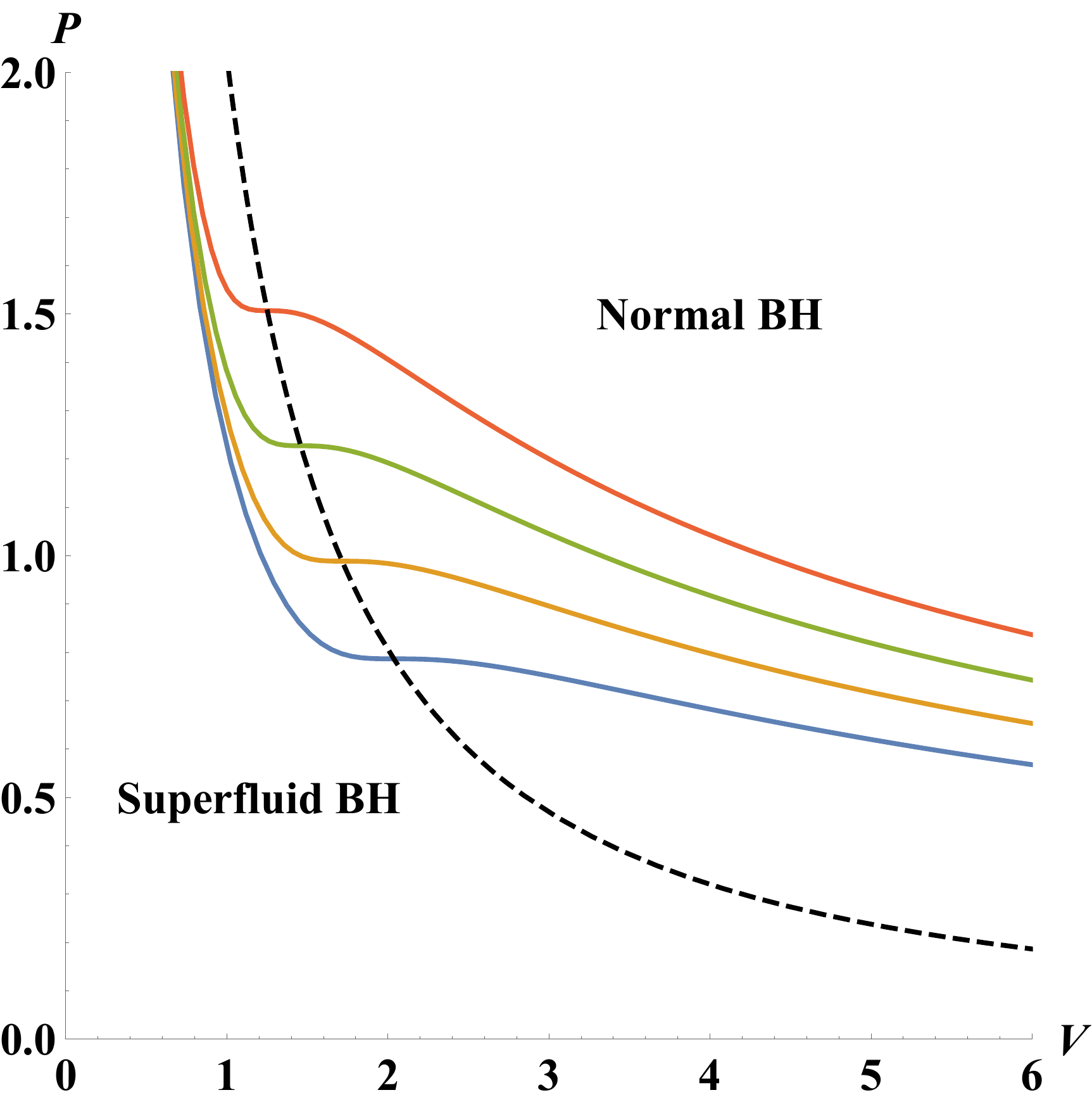}}
\hspace{0.2em}
\subfigure{\label{Superfluid_P_T}
\includegraphics[height=5cm]{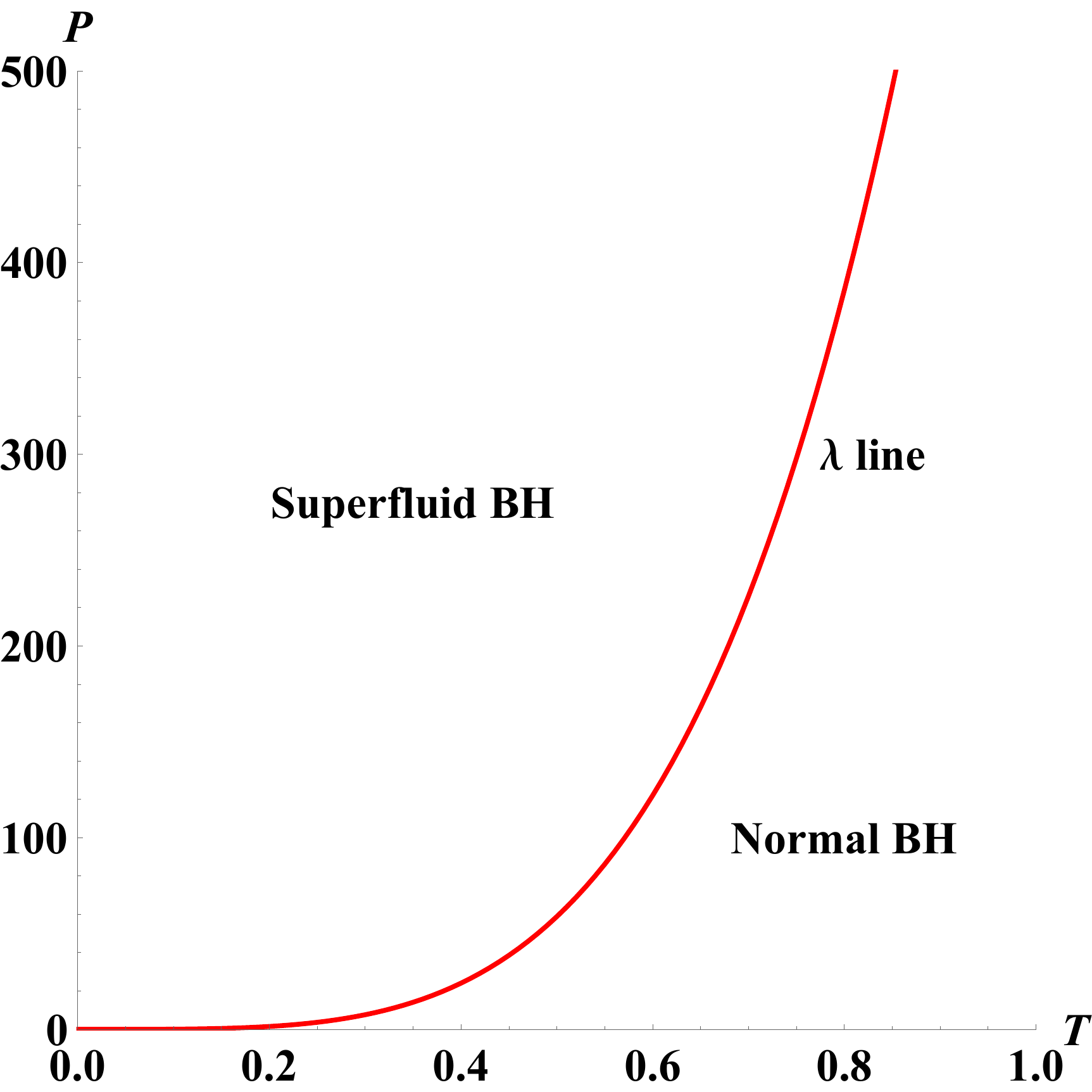}}
\hspace{0.2em}
\subfigure{\label{Superfluid_F_T}
\includegraphics[height=5cm]{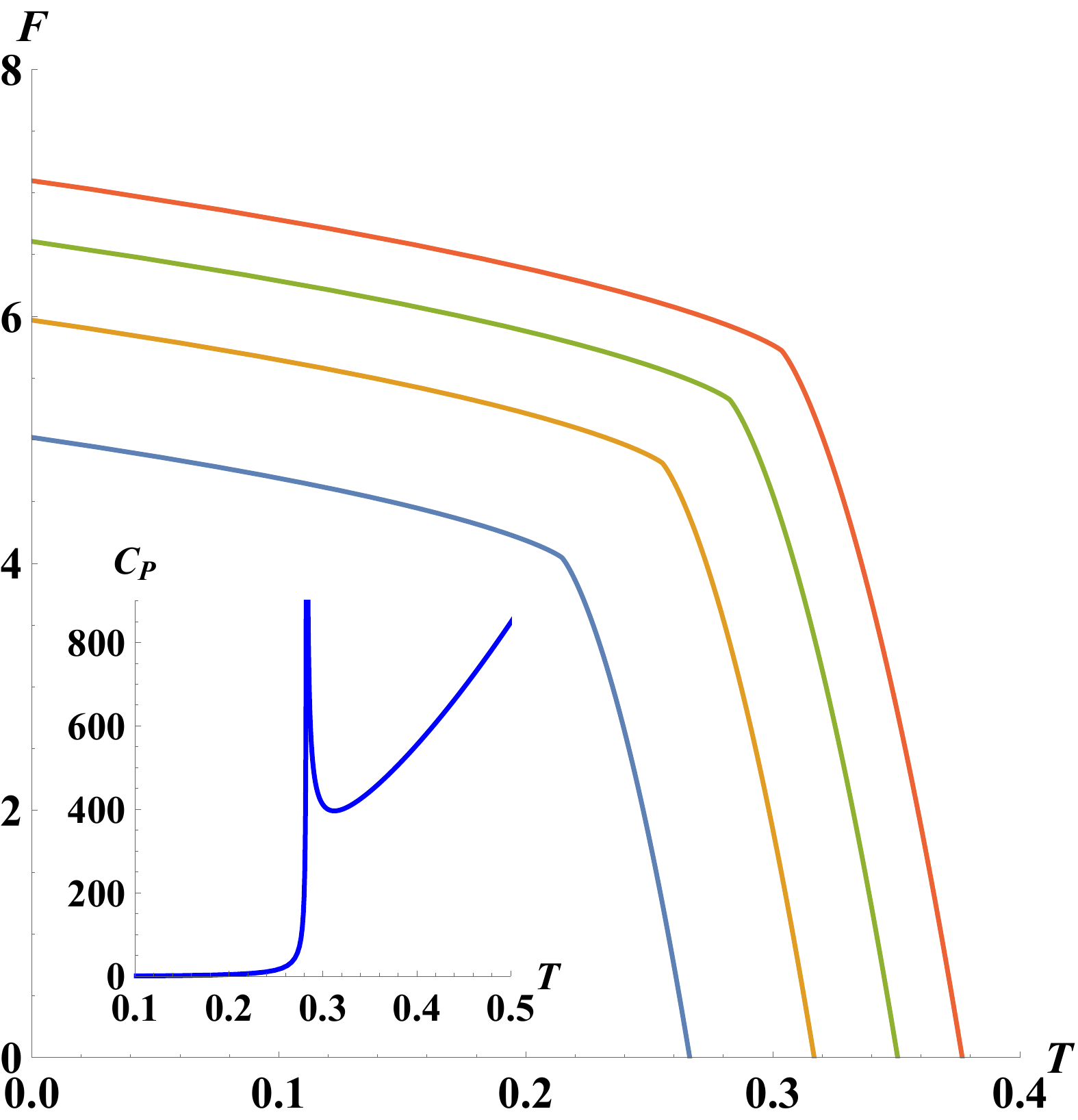}}}
\caption{\label{Superfluid} Superfluid $\lambda$ line phase transition for charged AdS black holes. In the $P-V$ diagram, the line of critical points is described as the the dashed line, and the solid lines are the isotherms with $T=0.17, 0.18, 0.19, 0.20$ from bottom to top. In the $P-T$ diagram, the line of critical points ($\lambda$ line) is shown as the red line. In the $F-T$ diagram, from bottom to top, the free energy curves are plotted with $P=2,4,6,8$, respectively. Setting $P=6$, the insert displays the typical behaviour of specific heat $C_P=- T \partial^2 F/\partial T^2$, which diverges at the critical temperature and shows a $\lambda$ structure. We have set $Q=1$ and $C=36$.}
\end{figure*}

Considering the $P-V$ criticality, the equation of state is described by $P=P(T,V)$ for fixed $(Q,C)$, with the thermodynamic critical points identified by
\begin{equation}
\left(\frac{\partial P}{\partial V}\right)_{T}=0, \quad\left(\frac{\partial^{2} P}{\partial V^{2}}\right)_{T}=0. \label{eq:con}
\end{equation}
Due to the complexity of Eqs. (\ref{eq:T}) and (\ref{eq:V}), there is no way to write an analytic equation of state $P(T,V)$. Detailed calculations with the chain rule show that the above condition can only be satisfied by $C=36 Q^2$ and
\begin{align}
P_c=\frac{3 Q^2}{8 \pi r^4_c}, \label{eq:critical_P}
\end{align}
where $P_c$ and $r_c$ denote the critical pressure and horizon radius, respectively. The critical volume and temperature are
\begin{align}
V_c &= \left(\frac{\pi}{3}\right)^{1/4} \left(\frac{C}{18 P_c}\right)^{3/4}, \label{eq:critical_V} \\
T_c &=\frac{1}{3} \left(\frac{2}{\pi }\right)^{3/4} \left(\frac{12 P_c}{C}\right)^{1/4}. \label{eq:critical_T}
\end{align}
As we can see, $V_c$ and $T_c$ are related to the values of $P_c$ and $C$. Thus for a fixed $C$ ($=36Q^2$), a line of critical points can be drawn in the $P-V$ or $P-T$ diagram, as shown in Fig. \ref{Superfluid}. In the $P-V$ diagram, each of the isotherms (solid lines) is a critical isotherm, and their inflection points constitute the line of critical points (dashed line) described by Eq. (\ref{eq:critical_V}), which corresponds to the one in the $P-T$ diagram described by Eq. (\ref{eq:critical_T}).

It is evident that at the line of critical points, second order (continuous) phase transitions occur due to the continuity of free energy $F=M-T S$ and the divergence of specific heat $C_P= - T \partial^2 F/\partial T^2 $ at critical temperatures (see the right panel in Fig. \ref{Superfluid}). This phenomenon is analogous to those that take place in condensed matter systems, such as normal fluid/superfluid transitions \cite{Gasparini:2008zz}, superconductivity \cite{Dasgupta:1981zz}, and paramagentism/ferromagnetism transitions \cite{Pathria:1996hda}. In particular, the shape of $C_P$ bears a strong resemblance to the one observed in the
fluid/superfluid transition of $^4\text{He}$, which is reminiscent of the symbol $\lambda$ \cite{Gasparini:2008zz}. It is then natural to interpret such a second order phase transition between small and large black holes as a normal fluid to superfluid phase transition.

For certain properties the
Bose-Einstein condensation (BEC) of interacting Bose gas can describe the superfluid transition of liquid $^4\text{He}$. Interestingly, we find that the phase diagrams presented in Fig. \ref{Superfluid} are very similar to the ones for the BEC of an imperfect Bose gas with hard-sphere potential \cite{huang1987} (see Appendix \ref{thermodynamics} for more details). Both the phase transition lines emanate from the origin, and extend to infinity in the $P-T$ diagrams. In the $P-V$ diagrams, both the phase transition lines diverge at zero volume, and monotonically decrease to zero at infinity volume. Moreover, we notice that the critical entropy for the BEC of hard-sphere Bose gas is uniquely determined by degrees of freedom (or particle number) of the system, i.e., $S_c \sim N$. A similar property holds for the critical entropy of black hole phase transitions, which is uniquely determined by degrees of freedom of the dual CFT, i.e., $S_c \sim C$. This can be easily seen by substituting Eq. (\ref{eq:critical_P}) into Eq. (\ref{eq:S}).

It would be interesting to study the critical phenomena near the second-order phase transition since they can reflect general properties of the system. Of primary interest near the $\lambda$ line are the critical behaviour of specific heat $C_{V} \propto |t|^{-\alpha}$ and correlation length $\xi \propto |t|^{-\nu}$, where $t \equiv (T-T_{\lambda})/T_{\lambda}$ \cite{Gasparini:2008zz}. From Eqs. (\ref{eq:S}-\ref{eq:V}), as well as the critical values given in Eqs. (\ref{eq:critical_P}-\ref{eq:critical_T}), we find
\begin{align}
& C_{V} \equiv T \left(\frac{\partial S}{\partial T}\right)_V \overset{T \rightarrow T_{\lambda}} {\longrightarrow} \pi C.
\end{align}
In other words, the specific heat has a finite limit at the $\lambda$ line and thus $\alpha = 0$. This result coincides with the calculations for the BEC of 3D hard-sphere gases \cite{huang1987}. For the critical exponent $\nu$, one feasible way to obtain an estimated value is by using the Josephson scaling law $d \nu = 2 - \alpha$, where $d$ is the spatial dimension \cite{Pathria:1996hda}. If we adopt the central dogma \cite{Almheiri:2020cfm} which states that as seen from an outside observer a black hole behaves like
a quantum system with $A/4G$ degrees of freedom, $d$ can be assumed to be $3$. This leads to $\nu = 2/3$, which is in agreement with analytical
result \cite{LeGuillou:1979ixc} and simulation \cite{Gruter:1997zz} for the BEC of 3D hard-sphere gas, as well as experimental data for liquid $^4\text{He}$ which gives $\nu=0.67$ \cite{Gasparini:2008zz}.

So far, it has been seen that the macroscopic phase transition of the black holes shows surprising similarities to the superfluid transition of liquid $^4\text{He}$, and especially, to the BEC of 3D hard-sphere gas. It is then natural to wonder if there are any similarities between their microscopic structures. To this end, the well-known Ruppeiner geometry \cite{Ruppeiner:1995zz} provides us an useful tool. In such a thermodynamic geometry, the fluctuation of entropy is associated with the line element in the Riemannian geometry
\begin{align}
\Delta l^{2}=-\frac{\partial^{2} S}{\partial x^{\mu} \partial x^{\nu}} \Delta x^{\mu} \Delta x^{\nu}, \label{eq:delta_S}
\end{align}
where $x^{\mu} = (U,Y^{i})$, $U$ is the internal energy and $Y^{i}$ are the other independent thermodynamic quantities. Tests on known models reveal that the scalar curvature $R$ obtained from Eq. (\ref{eq:delta_S}), can be an indicator for the microscopic interaction in a thermodynamic system. Concretely, the positive (negative) $R$ implies a
repulsive (attractive) interaction, whereas $R=0$ corresponds
to non-interaction. Note that the interaction here may not refer to the interparticle interaction, but rather a statistical interaction induced by the quantum statistics \cite{Janyszek1990}. For Bose–Einstein
statistics, the statistical interaction is attractive, while for Fermi–Dirac statistics it is repulsive.

\begin{figure}[t]
\centering
\includegraphics[height=6cm]{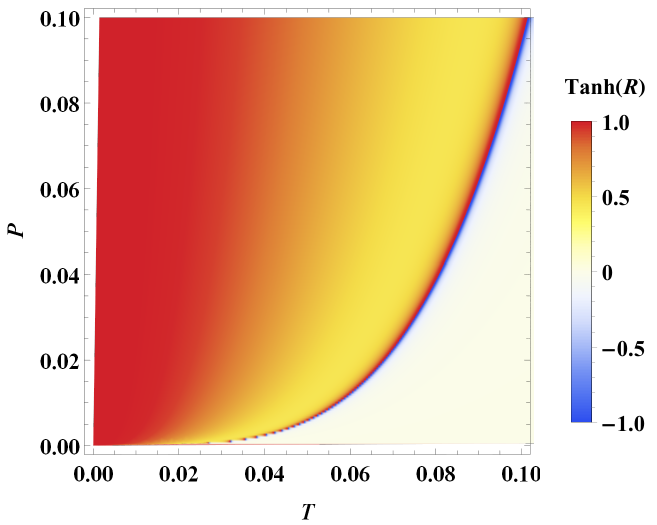}
\caption{\label{R_Superfluid} Behaviour of $R$ in the $P-T$ diagram.}
\end{figure}

Considering the existence of $V \delta P$ term in the first law (\ref{eq:first_mixed}), we would like to interpret $M$ as 
enthalpy for the gravitational system, and introduce an internal energy  $U=M-P V$ to compare with the standard thermodynamic system. For a fixed electric charge $Q$ and central charge $C$, choosing $(U, V)$ as fluctuation coordinates,
the line element in the $(S,P)$ plane can be reduced to \cite{Wei:2019uqg,Dehyadegari:2020ebz}
\begin{equation}
\label{eq:25}
\Delta l^{2}=\frac{1}{T}\left(\frac{\partial T}{\partial S}\right)_{P} \delta S^{2}-\frac{1}{T}\left(\frac{\partial V}{\partial P}\right)_{S} \delta P^{2}.
\end{equation}
The new thermodynamic volume $V$ is independent of entropy $S$, and thus the second term will not vanish, leading to a well-defined line element. Specially, for the charged AdS black hole, it is easy to see that $-(\partial_P \tilde{V})_{S}>0$ from Eqs. (\ref{eq:S}) and (\ref{eq:V}). By introducing the dimensionless parameter $s \equiv S/S_c$, we find that the expression for $R$ can be cast into a compact form,
\begin{align}
&R =\frac{64 \left(5 s^2+2 s+1\right) \left(3 s^4+12 s^3+1\right)}{\pi C (1-s)^3 \left(s^2+6 s+1\right)^2 \left(3 s^2+6 s-1\right)}.
\end{align}
It is clear that $R$ diverges at $s=1$, which corresponds to the $\lambda$ line in the $P-T$ diagram, as displayed in Fig. \ref{R_Superfluid}. Upon the $\lambda$ line, $R$ has a positive sign, while below it $R$ has a negative sign. This thus suggests an interesting microstructure for charged AdS black holes: the superfluid phase is dominated by repulsive interaction, while the normal phase is dominated by attractive interaction. 

Surprisingly, similar behaviour is also observed with the Ruppeiner scalar $R$ for 3D hard-sphere Bose gas (see Appendix \ref{Ruppeiner} for more details). Since the only interparticle interaction in the hard-sphere model is the repulsive interaction due to the particle collisions \cite{huang1987}, we can infer that attractive interaction in the gas phase originates from the statistical interaction induced by Bose–Einstein statistics. Based on the similarities between charged AdS black holes and 3D hard-sphere Bose gas, it is natural to conjecture that the attractive interaction in the black hole microstructure may arise from the statistical interaction, while the repulsive interaction comes from particle collisions.

Summarizing, we have demonstrated that there is a superfluid-like phase transition in the holographic
extended thermodynamics of charged AdS black holes. The associated phase behaviour shows surprising similarities
to the $\lambda$ transition of liquid $^4\text{He}$ and the BEC of 3D hard-sphere gas. We also find that the critical exponents share the same values $\alpha=0$ and $\nu = 2/3$. Besides the macroscopic thermodynamic properties, further investigations on the Ruppeiner geometry show that they also share a similar microscopic property: the superfluid (condensed) phase is dominated by a repulsive interaction, while the normal (gas) phase is dominated by an attractive interaction. These findings might cast new insights into the quantum aspect of black hole thermodynamics.

\begin{figure}[t]
\centering
\includegraphics[height=6cm]{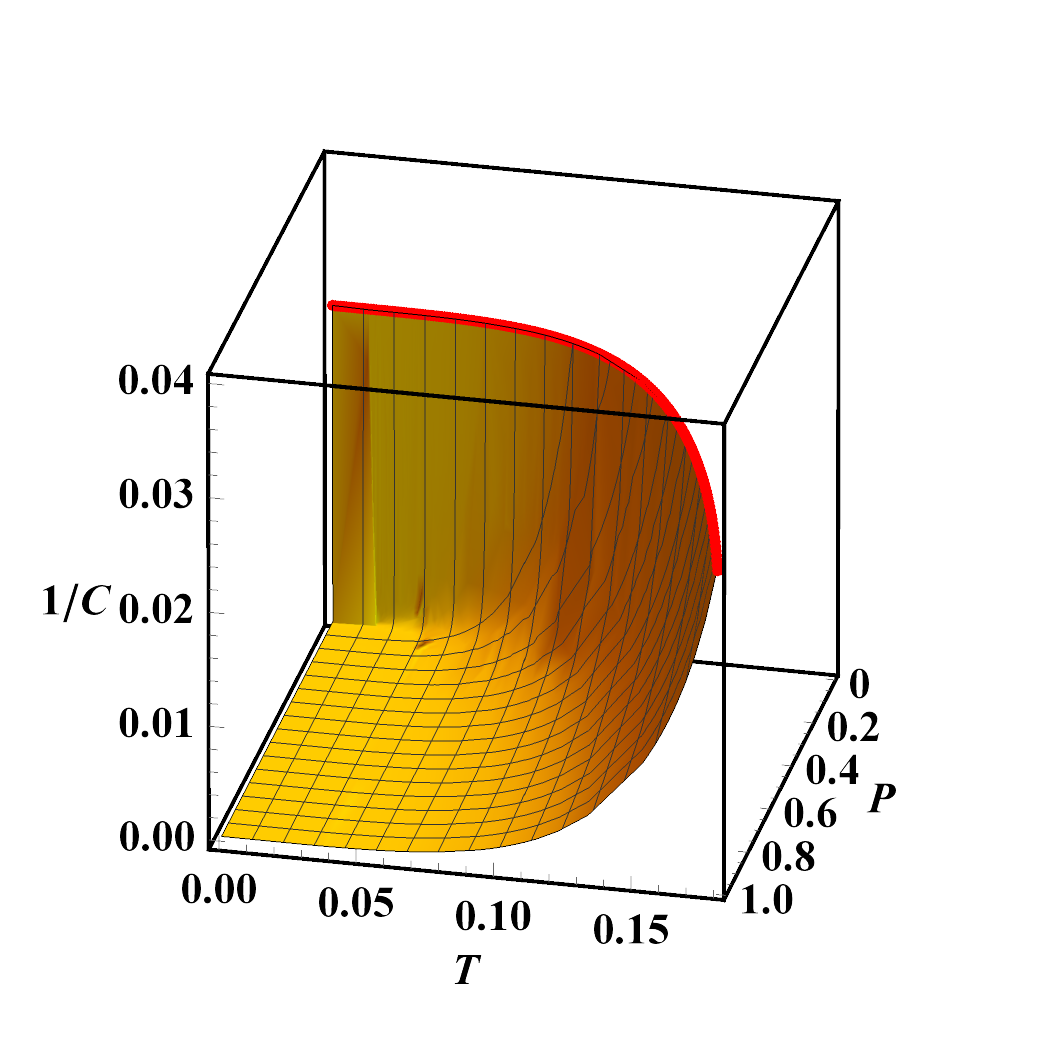}
\caption{\label{P_T_C} Phase transition in $(P,T,C)$ space. Here $Q=1$.}
\end{figure}

It is worth mentioning that there are some differences between the superfluid phase transition we discovered here and the one presented in Refs. \cite{Hennigar:2016xwd,Hennigar:2016ekz}. In the latter, the $\lambda$ line in $P-T$ diagram is a straight line with positive intercept, and the critical thermodynamic volume $V_c$ is a constant that independent of pressure $P$. In this work, the $\lambda$ line emanates from the origin, and the $V_c$ is $P$-related, as displayed in Fig. \ref{Superfluid}. Interestingly, both the $\lambda$ line and the $P-V_c$ curve bear a strong resemblance to those of 3D hard-sphere gas. In addition, the superfluid phase transition in Refs. \cite{Hennigar:2016xwd,Hennigar:2016ekz} occurs for hyperbolic black holes, while it occurs for spherical black holes here.

One may also wonder if there are any connections between the VdW-like phase transitions presented in Ref. \cite{Cong:2021fnf} and the superfluid-like phase transition we found here. To explore this question, one can consider an enlarged parameter space \cite{Hennigar:2016ekz}, as displayed in Fig. \ref{P_T_C}. The line of critical points (red), found in the $P-T$ plane with $C=36 Q^2$, represents a boundary where a surface of first-order phase transitions (yellow) terminates in the parameter
space $P-T-C$. For each constant $P$ slice, the critical point and the first-order phase transition form a VdW-like behaviour in the $1/C-T$ plane. 

Finally, we emphasize that the emergence of new phase behaviour arises from considering a variable $G$. On the one hand, it introduces a new thermodynamic volume $V$ for the black hole, see Eq. (\ref{eq:V}). On the other hand, it shows that both $G$ and $L$ are related to the thermodynamic pressure $P$ of the black hole, as described by Eq. (\ref{eq:G}). Consequently this results in  a new equation of state $P=P(T,V)$, leading to novel phase behaviour.

It would be interesting to reconsider the $P-V$ criticality of other AdS black holes, such as those with rotation, nonlinear electromagnetic field, and higher derivative curvature corrections, within the holographic framework that includes the variation of $G$, to see  if  superfluid-like phase transitions can still exist, or if some new phase behaviours are present. These investigations may reveal some new correspondences between ordinary thermodynamics and the thermodynamics of AdS black holes, in the context of gauge-gravity duality.

We are grateful to Robert B. Mann, Xiuming Zhang and Yan He for useful discussions and valuable comments. This work is supported by NSFC (Grant No. 12275183, 12275184,  12105191, and 12175212).

\appendix

\section{Thermodynamics for 3D hard-sphere Bose gas} \label{thermodynamics}
The primary material is derived from the computation presented in Ref. \cite{huang1987}. Considering a dilute system of $N$ identical spinless bosons of mass $m$, contained in a box of volume $V$, at very low temperatures. The bosons interact with one another through binary collisions characterized by the scattering length $a$ which is assumed to be positive. The corresponding Hamiltonian is given by
\begin{align}
\hat{H}=-\sum_{i} \frac{\hbar^{2}}{2 m} \nabla_{i}^{2}+\frac{4 \pi a \hbar^{2}}{m} \sum_{i<j} \delta\left(\boldsymbol{r}_{i}-\boldsymbol{r}_{j}\right).
\end{align}
 Treating the second term as a small perturbation, to the first order in $a$, the energy levels can be calculated as
\begin{align}
E_{n}=\sum_{\mathbf{p}} \frac{p^{2}}{2 m} n_{\mathbf{p}}+\frac{4 \pi a \hbar^{2}}{m V}\left(N^{2}-\frac{1}{2} \sum_{\mathbf{p}} n_{\mathbf{p}}^{2}\right), \label{eq:E}
\end{align}
where $n_{\mathbf{p}}$ is the number of bosons with momentum $\mathbf{p}$. Note that this expression is valid only under the conditions
\begin{align}
\frac{a}{v^{1/3}} \ll 1, \quad k a \ll 1,
\end{align}
where $v$ is the specific volume defined by $v=V/N$, and $k$ is the relative wave number of any pair of particles. Thus Eq. (\ref{eq:E}) becomes invalid if there are excited particles of high momentum. Especially, if we consider a very low temperature such that only a few particles are excited, the term $\frac{1}{2} \sum_{\mathbf{p}} n_{\mathbf{p}}^{2}$ can be approximated as $\frac{1}{2} n_0^2 $.

Let $n$ be an abbreviation for $\{n_{k}\}$, and let $\epsilon_n$ denote the first term of Eq. (\ref{eq:E}). Introducing the parameter $\xi \equiv n_0/N$, and referring the thermal wavelength as $\lambda = \sqrt{2 \pi \hbar^2 / m k_B T}$, the partition function can be written as
\begin{align}
Q_N &= e^{- \beta F_N} \nonumber \\
&=\sum_{n} e^{-\beta \epsilon_{n}} e^{-N\left(a \lambda^{2} / v\right)\left(2-\xi^{2}\right)} \nonumber \\
&=Q_{N}^{(0)}\left\langle e^{-N\left(a \lambda^{2} / v\right)\left(2-\xi^{2}\right)}\right\rangle_{0},
\end{align}
where $F_N \equiv F_N(T,V,N)$ is the Helmholtz free energy of the system. The  $Q_{N}^{(0)}=\sum_{n} e^{-\beta \epsilon_{n}}$ is the partition function of the ideal Bose gas, and $\left\langle \cdots \right\rangle_{0}$ represents the thermodynamic average with respect to the ideal Bose gas. For the calculation of the partition function, the following regions are considered:
\begin{equation}
a/\lambda \ll 1, \quad a \lambda^2/ v \ll 1,  
\end{equation}
since they are the only dimensionless parameters in the problem involving $a$.

\begin{figure*}[t]
\center{\subfigure{\label{Bose_P_v}
\includegraphics[height=6cm]{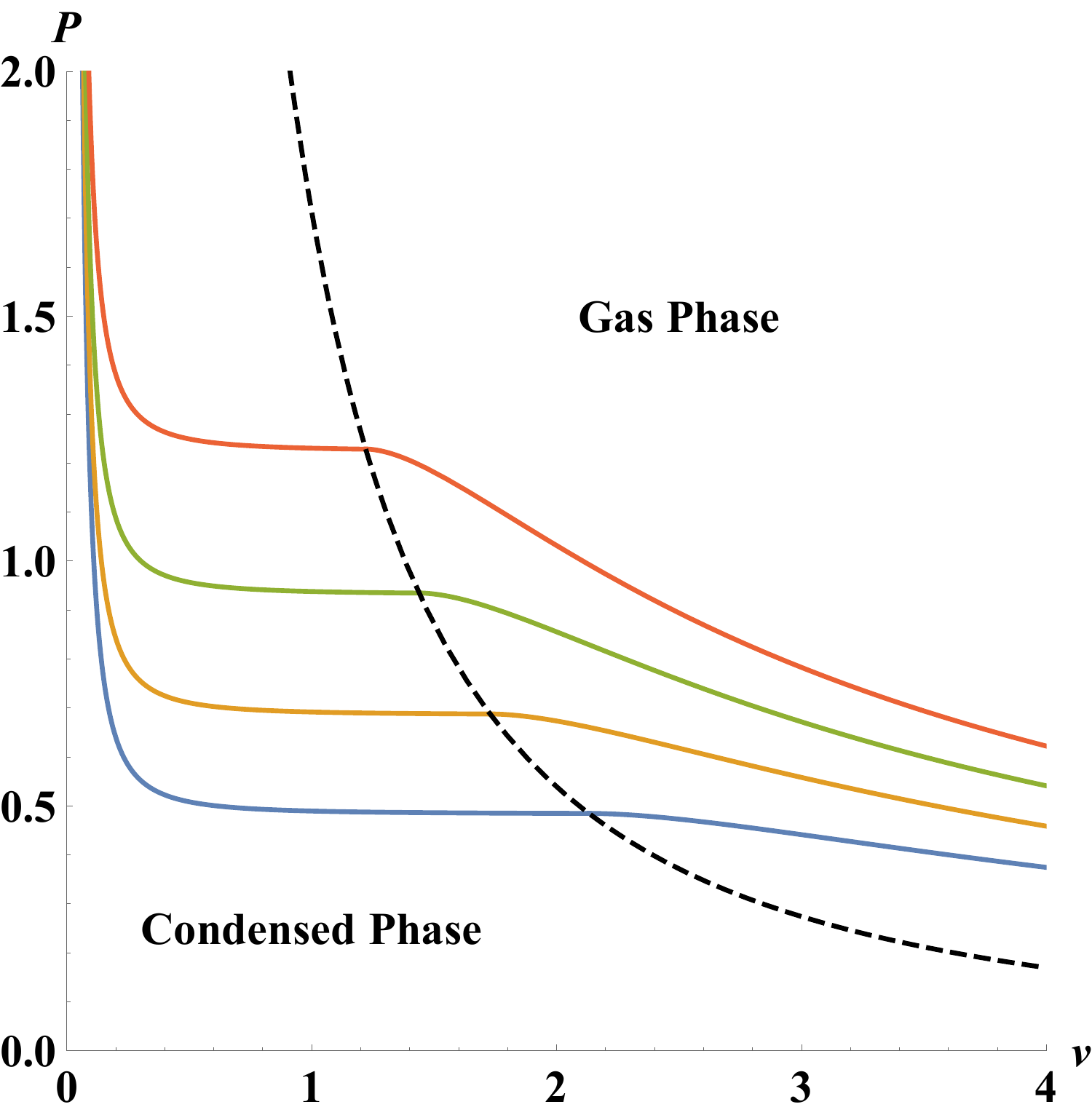}}
\hspace{1em}
\subfigure{\label{Bose_P_T}
\includegraphics[height=6cm]{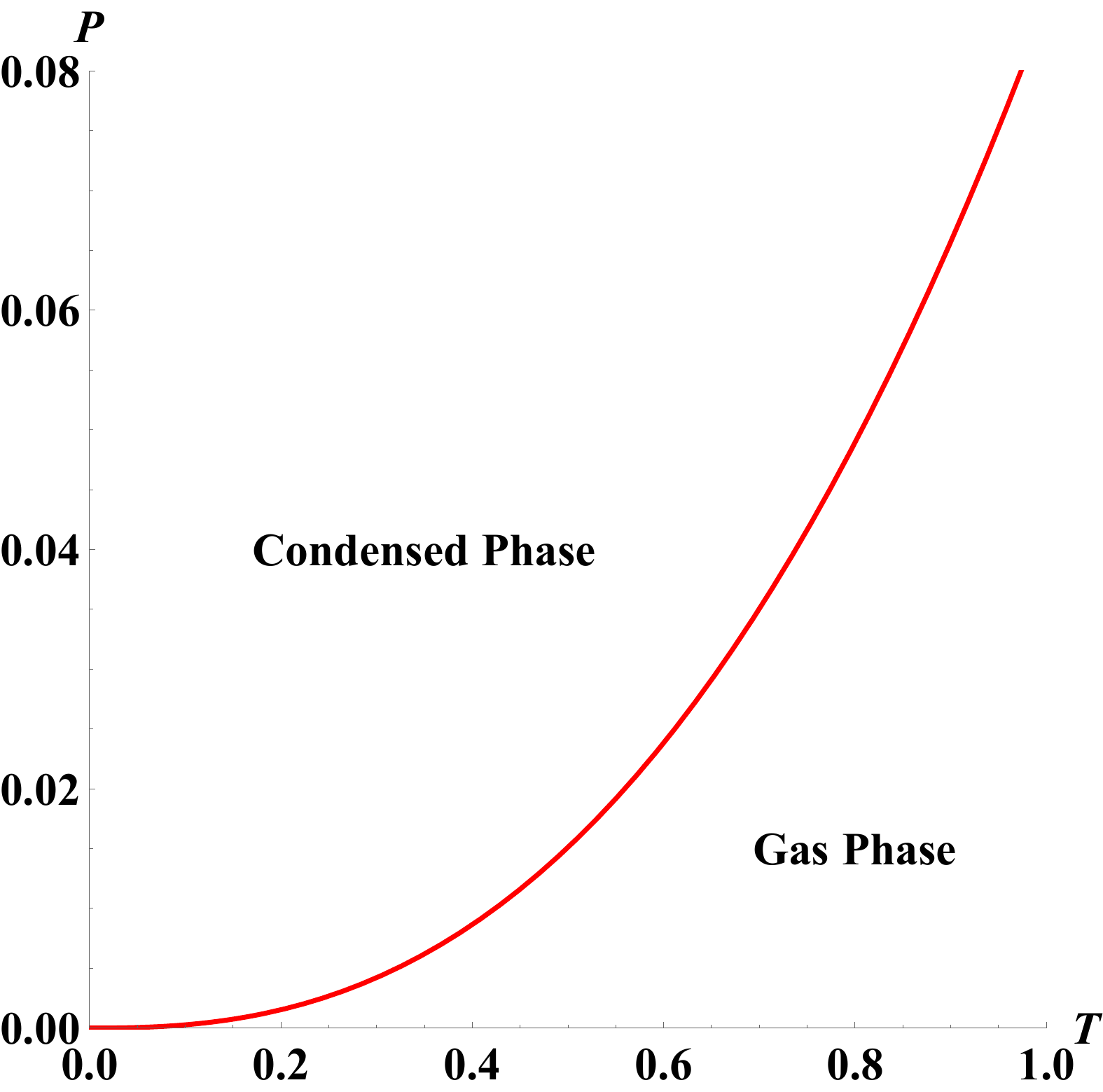}}}
\caption{\label{Bose} Bose-Einstein condensation for imperfect Bose gas with hard-sphere potential. In the $P-v$ diagram, the dashed line is the transition line, and the solid lines are the isotherms with $T=0.2, 0.23, 0.26, 0.29$ from bottom to top. In the $P-T$ diagram, the transition line is denoted as the red line. We have set $m=\hbar=k_B=1$ and $a=0.001$.}
\end{figure*}

Hence the free energy per particle can be calculated as
\begin{align}
\frac{F_N}{N} &= \frac{F_N^{(0)}}{N} - \frac{k_B T}{N} \ln \left\langle e^{-N\left(a \lambda^{2} / v\right)\left(2-\xi^{2}\right)}\right\rangle_{0}\nonumber \\
& \approx \frac{F_N^{(0)}}{N} + k_B T \frac{a \lambda^2}{v} \left\langle \left(2-\xi^2\right) \right\rangle_{0} \nonumber \\
& = \frac{F_N^{(0)}}{N} + \frac{4 \pi a \hbar^2}{m v} \left(1-\frac{1}{2}\overline{\xi^{2}} \right). \label{eq:free}
\end{align}
Here $ \left\langle \xi^2 \right\rangle_{0} \equiv \overline{\xi^{2}}$. Considering that for ideal Bose gas we have
\begin{align}
\left\langle n_{k}^2 \right\rangle_{0} -\left\langle n_{k} \right\rangle^2_{0} = \left\langle n_{k} \right\rangle_{0},
\end{align}
and thus
\begin{align}
\overline{\xi^{2}} -\overline{\xi}^2 = \overline{\xi}/N.
\end{align}
For large $N$, $\overline{\xi^{2}} \approx \overline{\xi}^2$, and Eq. (\ref{eq:free}) can be further expressed as
\begin{align}
\frac{F_N}{N} \approx \frac{F_N^{(0)}}{N} + \frac{4 \pi a \hbar^2}{m v} \left(1-\frac{1}{2} \overline{\xi}^2 \right).
\end{align}
The pressure can be then calculated as
\begin{align}
P &= -\left(\frac{\partial F_N}{\partial v} \right)_{T, N} \nonumber \\
&=P^{(0)} + \frac{4 \pi a \hbar^2}{m} \left[\frac{1}{v^2} \left(1-\frac{1}{2} \overline{\xi}^2 \right) + \frac{\overline{\xi}}{v} \frac{\partial \overline{\xi}}{\partial v} \right],
\end{align}
where $P^{(0)}$ is the pressure of the ideal Bose gas. Utilizing the expression for $\overline{\xi}$,
\begin{align}
\overline{\xi}=\left\{\begin{array}{ll}
0 & \left(\lambda^3/v \leq g_{3/2}(1)\right) \\
1-v/v_c=1- \left(T/T_c\right)^{3/2} & \left(\lambda^3/v \geq g_{3/2}(1)\right),
\end{array}\right.
\end{align}
where $g_{n} (z) = \sum\limits_{l=1}^{\infty} z^{l}/l^{n}$ are the Bose-Einstein functions, and especially, 
\begin{align}
g_{3/2} (1) \approx 2.612, \quad g_{5/2} (1) \approx 1.342.
\end{align}
The critical values $v_c$ and $T_c$ can be identified by using $\lambda^3/v = g_{3/2}(1)$.

The equation of state for gas phase ($v>v_{c}, T>T_{c}$) and condensed phase ($v<v_{c}, T<T_{c}$) can be then calculated as
\begin{align}
P=\left\{\begin{array}{ll}
P^{(0)}+ \frac{4 \pi a \hbar^2}{m v^2} & \left(v>v_{c}, T>T_{c}\right) \\
P^{(0)}+\frac{2 \pi a \hbar^{2}}{m}\left(\frac{1}{v^{2}}+\frac{1}{v_{c}^{2}}\right) & \left(v<v_{c}, T<T_{c}\right).
\end{array}\right.
\end{align}
with
\begin{align}
P^{(0)}=\left\{\begin{array}{ll}
\frac{k_B T}{v} \sum\limits_{l=1}^{\infty} b_{l} \left(\frac{\lambda^3}{v}\right)^{l-1} & \left(v>v_{c}, T>T_{c}\right) \\
g_{5/2}(1) \frac{k_B T}{\lambda^3} & \left(v<v_{c}, T<T_{c}\right).
\end{array}\right.
\end{align}
The $b_l$ are the virial coefficients, and turn out to be
\begin{align}
&b_1=1, \quad b_2=-\frac{1}{4 \sqrt{2}} \approx -0.17678, \nonumber \\
&b_3=\frac{1}{8}-\frac{2}{9 \sqrt{3}} \approx -0.00330,
\end{align}
and so on. The associated isotherm and $P-T$ diagram are displayed in Fig. \ref{Bose}. 

The entropy for the system is given by
\begin{align}
S&=-\left(\frac{\partial F_N}{\partial T}\right)_{N, V} \nonumber \\
 &=\left\{\begin{array}{ll}
k_B N \left[\frac{5}{2} \frac{v}{\lambda^{3}} g_{5 / 2}(z)-\log z\right] & \left(v>v_{c}, T>T_{c}\right) \\
k_B N \left[ \frac{5}{2} \frac{v}{\lambda^{3}} g_{5 / 2}(1)\right] & \left(v<v_{c}, T<T_{c}\right).
\end{array}\right.
\end{align}
As we can see, the critical entropy $S_c$ is uniquely determined by the particle number $N$, i.e., $Sc \sim N$.

The specific heat for the gas phase can be calculated as
\begin{align}
C_{V} & \equiv - T \left( \frac{\partial^2 F_N}{\partial T^2}\right)_{N,V} \nonumber \\
& =\frac{3}{2} k_B N \sum_{l=1}^{\infty} \frac{5-3 l}{2} b_{l}\left(\frac{\lambda^{3}}{v}\right)^{l-1} \nonumber \\
& =\frac{3}{2} k_B N \Bigl[1+0.0884\left(\frac{\lambda^{3}}{v}\right)+0.0066\left(\frac{\lambda^{3}}{v}\right)^{2}+\cdots\Bigr].
\end{align}
Similarly, the specific heat for the condensed phase is given by
\begin{align}
C_{V} = \frac{15}{4} k_B N g_{5/2}(1) \frac{v}{\lambda^3} + 3 g_{3/2}(1) \frac{ a \left(4 \lambda^3_c-\lambda ^3\right)}{2 \lambda ^4}.
\end{align}
At the transition point, the specific heat has a finite jump, 
\begin{align}
\frac{\Delta C_V}{N k_B} = \frac{9 a}{2 \lambda_c} g_{3/2} (1),
\end{align}
and thus we can say that the Bose-Einstein condensation here is a second-order transition. The finite jump also indicates that the critical exponent $\alpha$, governing the scaling behaviour of specific heat $C_{V} \propto |t|^{-\alpha}$ ($t \equiv (T-T_{c})/T_{c}$), takes $\alpha=0$. Another critical exponent of interest to us describes the scaling behaviour of correlation length $\xi \propto |t|^{-\nu}$, which can be inferred by using the Josephson scaling law $d \nu = 2 - \alpha$, where $d$ is the spatial dimension. Taking $d=3$, $\nu=2/3$ is obtained.

\section{Ruppeiner geometry for 3D hard-sphere Bose gas} \label{Ruppeiner}

\begin{figure}[t]
\centering
\includegraphics[height=6cm]{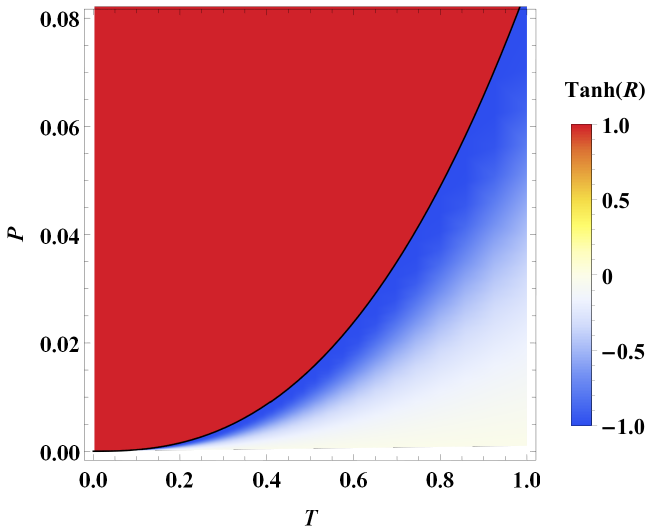}
\caption{\label{R_Bose} Behaviour of $R$ in the $P-T$ diagram. Here $a=0.001$.}
\end{figure}

To conveniently calculate the Ruppeiner scalar $R$, the line element can be chosen as \cite{Wei:2019uqg}
\begin{equation}
\label{eq:25}
\Delta l^{2}=\frac{C_V}{T^2} \delta T^{2}- \frac{(\partial_V P)_{T}}{T} \delta V^{2}.
\end{equation}
After simple calculation, the $R$ for condensed phase is given by (setting $m=\hbar=k_B=1$ hereafter)
\begin{equation}
R=\frac{v}{2 a \lambda ^2 N} +\frac{10 \lambda ^3+3 v g_{3/2} (1)}{30 v g_{5/2} (1) N} + \mathcal{O} ( (a/\lambda)^1).
\end{equation}
The $R$ for gas phase, considering the leading three terms of the virial expansion for state equation and specific heat, is calculated as
\begin{equation}
R=\frac{972 \sqrt{2} \lambda ^3 v^3 Z}{N X^2 Y^2} + \mathcal{O} ( (a/\lambda)^1),
\end{equation}
where
\begin{align}
X&= \left(27-16 \sqrt{3}\right) \lambda ^6 -18 \sqrt{2} \lambda ^3 v +72 v^2, \\
Y &= 4 \left(16 \sqrt{3}-27\right) \lambda ^6 +27 \sqrt{2} \lambda ^3 v +432 v^2, \\
Z &= 10 A \lambda ^{12} + B \lambda ^9 v -48 C \lambda ^6 v^2 -72 D \lambda ^3 v^3-311040 v^4,
\end{align}
and
\begin{align}
A&=288 \sqrt{3}-499, \quad B=83485 \sqrt{2}-48240 \sqrt{6}, \nonumber \\
C&=3645-2160 \sqrt{3}, \quad D=5120 \sqrt{6}-9045 \sqrt{2}.
\end{align}

Taking $a=0.001$ for example, the behaviour of $R$ is displayed in Fig. \ref{R_Bose}. The black line denotes the phase transition line. As we can see, $R$ goes to infinity at the transition line, which can be a representation of Bose-Einstein condensation. Below the the transition line, $R$ has negative sign, which suggests that the gas phase is dominated by the attractive interaction. Upon the transition line, $R$ has positive sign, implying that the condensed phase is dominated by the repulsive interaction. Since the only interparticle interaction in the hard-sphere model is the repulsive interaction due to the particle collisions, we can infer that attractive interaction in the gas phase originates from the statistical interaction induced by Bose–Einstein statistics.

\end{document}